\documentclass[useAMS,usenatbib]{mn2e}
\usepackage{epsfig}

\def\aap{A\&A}
\def\araa{ARA\&A}
\def\mnras{MNRAS}
\def\aj{AJ}
\def\apj{ApJ}
\def\pasp{PASP}

\def\cygx2{Cyg~X-2}

\def\spose#1{\hbox to 0pt{#1\hss}}

\def\um{$\mu$m}

\def\fw{{\rm W~cm}$^{-2} \mu$m$^{-1}$\ } 
\def\lta{\mathrel{\spose{\lower 3pt\hbox{$\mathchar"218$}}
     \raise 2.0pt\hbox{$\mathchar"13C$}}}
\def\gta{\mathrel{\spose{\lower 3pt\hbox{$\mathchar"218$}}
     \raise 2.0pt\hbox{$\mathchar"13E$}}}

\newcommand{\eg}{{\it e.g.,}}
\newcommand{\cf}{{\it c.f.,}}

\title{Excess mid-IR Emission in Cataclysmic Variables}

\author[G. Dubus et al.]{G. Dubus$^{1,2,3}$\thanks{Present address: Laboratoire Leprince-Ringuet}
, R. Campbell$^{4}$, B. Kern$^{1}$, R. E. Taam$^{5}$ and H. C. Spruit$^{6}$.\\
$^{1}$California Institute of Technology, Pasadena, CA 91107, USA\\
$^{2}$Laboratoire Leprince-Ringuet, CNRS/IN2P3, Ecole Polytechnique, F-91128, Palaiseau, France\\
$^{3}$Institut d'Astrophysique de Paris, 98bis boulevard Arago, F-75012, Paris, France\\
$^{4}$W. M. Keck Observatory, 65-1120 Mamalahoa Highway, Kamuela, HI 96743, USA\\
$^{5}$Department of Physics \& Astronomy, Northwestern University, Evanston, IL 60208, USA\\
$^{6}$Max-Planck-Institut f\"ur Astrophysik, Postfach 1317, D-85741 Garching, Germany}

\begin{document}
\date{Draft \today}
\maketitle

\begin{abstract}
We present a search for excess mid-IR emission due to circumbinary
material in the orbital plane of cataclysmic variables (CVs). Our
motivation stems from the fact that the strong braking exerted by a
circumbinary (CB) disc on the binary system could explain several
puzzles in our current understanding of CV evolution. Since
theoretical estimates predict that the emission from a CB disc can
dominate the spectral energy distribution (SED) of the system at
$\lambda >$ 5~\um, we obtained simultaneous visible to mid-IR SEDs for
eight systems. We report detections of SS~Cyg at 11.7~\um\ and AE~Aqr
at 17.6~\um, both in excess of the contribution from the secondary
star. In AE~Aqr, the IR likely originates from synchrotron-emitting
clouds propelled by the white dwarf. In SS~Cyg, we argue that the
observed mid-IR variability is difficult to reconcile with simple
models of CB discs and we consider free-free emission from a wind. In
the other systems, our mid-IR upper limits place strong constraints on
the maximum temperature of a putative CB disc. The results show that
if any sizeable CB disc are present in these systems, they must be
self-shadowed or perhaps dust-free, with the peak thermal emission
shifted to far-IR wavelengths.
\end{abstract}

\begin{keywords}
binaries: close -- stars: evolution -- stars: cataclysmic variables -- stars: individual (QR~And, RX~And, AE~Aqr, HU~Aqr, SS~Cyg, IP~Peg, WZ~Sge, RW~Tri)
\end{keywords}

\section{INTRODUCTION}
Studies of cataclysmic variables (CVs) have largely focused on
wavelengths $\lambda \lta$ 1~\um\ where accretion processes dominate
the system emission. Surprisingly little is known about the IR
properties of CVs \citep{dhillonconf}. Near-IR observations show
contributions above that of the K/M donor star spectrum which peaks
around $\lambda\sim 3~\mu$m. In most cases colors are bluer than
expected because of residual light from the accretion disc. In others
the colors are redder, which is interpreted as the emission from the
evolved secondary components or from the cool outer regions of
unusually large accretion discs \citep{dhillon,2mass}. In well-known
systems, the spectral energy distribution (SED) can be decomposed into
its different components \citep[\eg][]{ciardi}. \citet{harrison} then
argued that U~Gem and SS~Cyg have excess near-IR emission above that
expected from the secondary star.

A possible source of additional IR emission is cold material located
outside the binary system. Some CVs have P~Cyg lines formed in an
accretion disc wind ($v\approx 4000$~km~s$^{-1}$, $\dot{M}_{\rm
wind}\approx 10^{-10}$~M$_\odot$~yr$^{-1}$); rapidly spinning white
dwarf magnetic fields can propel large fractions of the gas out of the
system (\eg\ AE~Aqr); and runaway nuclear burning in novae leads to
the ejection of the white dwarf envelope ($\sim 10^{-4}$~M$_\odot$). A
small fraction of this ejected material may remain bound to the binary
and form a {\em circumbinary disc} (CB disc). Another possible source
of material is residual gas from the common-envelope evolution phase
which precedes the onset of mass transfer. The inner edge of a CB disc
is tidally coupled to the binary system, leading to the extraction of
angular momentum from the orbital motion \citep{webbink}. The CB disc
therefore expands as angular momentum is redistributed by viscous
interactions \citep{pringle} resulting in a very efficient orbital
braking mechanism if the CB disc is sufficiently dense
\citep{spruittaam,taam}.

Recently, it has been pointed out by \citet{andronov} that the angular
momentum loss associated with magnetic braking based on the
\citet{skumanich} law severely underestimates the rotational velocities
of single stars in open clusters, implying that the magnetic braking
that has been the basis for CV evolution \citep[\eg\ ][]{vz} is
significantly reduced \citep[\cf\ ][]{ivanova}. Hence, additional
angular momentum losses are required to understand the diversity and
level of the mass transfer rates in CVs. CB discs can provide this
angular momentum loss (in addition to gravitational wave radiation and
magnetic braking) that may help to solve some of the puzzles of CV
evolution, such as the lack of a peak at the minimum in the orbital
period distribution, the order of magnitude spread in mass transfer
rates $\dot{M}_{\rm t}$ at a given $P_{\rm orb}$ or the very high
$\dot{M}_{\rm t}$ implied by steady nuclear burning on the white dwarf
in the supersoft sources (see, for example, \citealt{taamsd}). As mass
in the CB disc builds up, a feedback effect increases $\dot{M}_{\rm
t}$ above the values expected from standard evolution eventually
leading to the dissolution of the binary \citep{taam,dubus,taamsd}. At
this point, CB discs will typically be optically thick with sizes
$\gta$1-10~AU and effective temperatures of a few 1000~K in their
innermost regions. These properties make them very similar to
protostellar discs.

Observations of stationnary, faint lines with narrow velocity widths
in several objects suggest the presence of circumbinary material
(Z~Cam, \citealt{robinson}; IP~Peg, \citealt{piche}; AM~CVn,
\citealt{solheim}; SS~Cyg, \citealt{steeghs}; QR And,
\citealt{deufel}; BY~Cam, \citealt{mouchet}). The continuum emission
from a CB disc (possibly responsible for the narrow lines) could
easily have escaped detection up until now.  Spectral energy
distributions (SEDs) calculated by \citet{dubus} and \citet{taamsd}
show that massive CB discs start dominating over the binary emission
at about 5$\mu$m in mid-IR. The mid- and far-IR remain unexplored,
with {\em IRAS} providing loose upper limits or detections subject to
source confusion (see below).

Motivated by the prospect of finding excess emission from a CB disc in
the poorly known mid-IR regime, we observed a total of eight systems
(Tab.~\ref{tab:obj}). A SED spanning the visible and mid-IR is crucial
to identify the upturn caused by the CB disc, and we observed these
systems nearly simultaneously with the Palomar 60-inch (visible) and
Keck I telescopes (near- and mid-IR). Our prime objective was SS~Cyg
because of its brightness, parallax distance and hint of excess IR
emission \citep{harrison}. The SED of CVs involves various, poorly
constrained components (accretion emission, white dwarf, secondary
etc). Their contribution is {\em a priori} expected to decrease at
mid-IR wavelengths, making any excess more straightforward to identify
than at visible or near-IR wavelengths. However, we caution that since
very little is known about the mid-IR emission of CVs, any excess will
not necessarily imply circumbinary material.

The other targets are well-studied systems representing a wide range
of CV sub-classes as we have little {\em a priori} knowledge of which
types of CVs are more likely to have more circumbinary material. The
list includes binaries with $P_{\rm orb}$ below the period gap, driven
by gravitational wave radiation, and some with $P_{\rm orb}> 3$~hrs
presumably driven by magnetic wind braking. We included the supersoft
QR~And despite its faintness and large distance to test the
possibility that a large CB disc drives the high mass transfer rate.

Some of the CVs had previously published mid-IR measurements. The
earliest mid-IR measurements of CVs are those of \citet{bsc} who
reported ground-based 2$\sigma$ upper limits at 10~\um\ of $<$27~mJy
for SS~Cyg and $<$36~mJy for AE~Aqr. The more recent {\em ISO}
measurements of AE~Aqr by \cite{abada2} are discussed in
\S\ref{sync}. All other published measurements are based on {\em
IRAS}. \citet{jameson} report detecting SS~Cyg as it underwent a dwarf
novae outburst. The fluxes at 12 and 25~\um\ are $\sim$55 and 30~mJy
(the quiescent levels are $<$10 and $<$8~mJy). But a later study using
the same data \citep{iras} quotes a 3$\sigma$ upper limit of $<$30~mJy
at 12~\um\ and a 150$\pm$40~mJy detection at 25~\um\ for SS~Cyg
($<$110~mJy at 60~\um). \citet{iras} also report 3$\sigma$ upper
limits for RX~And of $<$30~mJy at 12 and 25~\um. Source confusion,
particularly in the galactic plane, is a major issue with {\em IRAS},
which could explain some of the discrepancies\footnote{For instance,
the referee points out a possible confusion between SS~Cyg and the
star Misselt 1 which is $\sim$ 2\arcmin\ away.} and very high far-IR
fluxes \citep{iras2}. At 12~\um\, the angular resolution is only
$\sim$30\arcsec. There is much that remains to be clarified as regards
the mid-IR emission from CVs.

The paper is organized as follows. In \S2 we describe our Palomar and
Keck observational setup and the data reduction. The visible to
near-IR SEDs of the systems are shown in \S3 where we fit the expected
secondary star SEDs to the observations and investigate any anomalous
excess.  Only two systems are detected at mid-IR wavelengths: AE~Aqr
and SS~Cyg. The measurements and consequences on models are set out in
\S4 (AE~Aqr) and \S5 (SS~Cyg). In \S6 (which may be skipped by the
observationally minded reader) we return to our original motivation for
this project and explore the implications of our mid-IR upper limits
on the presence and observational prospects for detecting CB discs. We
conclude in \S7.

\begin{table}
\caption{Observed cataclysmic variables.}
\begin{center}
\label{tab:obj}
\begin{tabular}{@{}lrrrrr}
\hline
Object & Clas. & $P_{\rm orb}$ & Sec. & Dist. & Refs\\
 & & {\small (hrs)} & & {\small (pc)} & \\
\hline
AE Aqr & IP    & 9.88 & K4V   & 102$^p$ & (1)\\
SS Cyg & DN    & 6.60 & K4V   & 159$^p$ & (2)\\
RX And & DN    & 5.04 & K5V   & 191$^w$ & (3)\\
HU Aqr & P     & 2.08 & M4V   & 191$^b$ & (4)\\
IP Peg & DN    & 3.80 & M4V   & $>$130$^b$ & (5)\\
WZ Sge & DN/IP & 1.26 & $>$M7V & 48$^w$ & (6)\\
RW Tri & NL    & 5.57 & M0V   & 341$^p$ & (7)\\
QR And & SS    & 15.8 & ?    & $>$2000$^c$ & (8)\\
\hline
\end{tabular}
\end{center}

\medskip
Spectral types determined from visible or IR spectroscopy. Distances
determined from: parallax ($p$), white dwarf UV emission models ($w$),
$K$-band flux \citep[][$b$]{bailey}, Ca{\sc ii} line ($c$). Data from
\citealt{ritter} and (1) \citealt{friedjung}; (2) \citealt{harrison};
(3) \citealt{smith,sepinsky}; (4) \citealt{glenn}; (5)
\citealt{szkody} (6) \citealt{ciardi,smak} (7)
\citealt{dhillon,mcarthur}; (8) \citealt{beuermann}.
\end{table}

\section{OBSERVATIONS AND REDUCTION}

The observations were carried out on the nights of September 15 and
16, 2002 using the 10-m Keck I telescope at IR wavelengths and the
Palomar 60-inch for visible light. The telescope pointings were
time-correlated to obtain in as much as possible near simultaneous
photometry across the spectrum. Some additional mid-IR data was
obtained at Keck I using director's time on August 18,
2002.

\begin{table*}
\caption{Average measured flux density $S_\nu$ in mJy. \label{tab:obs}}
\begin{minipage}{150mm}
\begin{tabular}{@{}llrrrrrrrrrrr}
\hline
 & &$U$ & $B$ & $V$ & $R$ & $I$ & $J$ & $H$ & $K$ & $M$ & SiC & 17.6\\
\hline
 AE Aqr &  &  22.2 &  55.5 & 104.3 & 156.1 & 211.5 & 274.6 & 301.0 & 215.3 &  55.3$\pm$2.0 &  24.1$\pm$1.0 &  26.8$\pm$5.5 \\
 SS Cyg &  &  26.7 &  29.9 &  47.7 &  73.3 & 104.5 & 130.0 & 146.9 & 110.7 &  33.2$\pm$1.5 &  11.6$\pm$0.9 &  3.2$\pm$6.7 \\
 RX And  & &  13.7 &   9.0 &  9.7 &  11.7 &  16.3 &  19.4 &  21.3 &  16.8 &  $<$14.5 &   $<$6.1 & $<$152.0 \\
 QR And  & &  60.4 &  47.3 &  40.8 &  36.7 &  32.6 &  20.1 &  14.2 &  10.3 &   $<$6.4 &   $<$4.0 & ...\\
 HU Aqr &  &   0.7 &   0.5 &   0.6 &   0.2 &   0.8 &   1.2 &   1.3 &   1.0 &   $<$6.6 &   $<$5.3 & ...\\
 IP Peg &  &   0.5 &   1.8 &   1.4 &   1.2 &   4.6 &  11.2 &  12.6 &  10.3 &   $<$8.4 &  $<$10.5 & ...\\
 WZ Sge &  &   5.3 &   4.3 &   3.9 &   3.8 &   3.4 &   2.3 &   2.3 &   2.4 &   $<$7.6 &   $<$4.7 & ...\\
 RW Tri &  &  18.6 &  21.4 &  22.1 &  23.5 &  26.5 &  23.3 &  21.9 &  15.7 &   $<$7.3 &   $<$3.0 & ...\\
\hline
\end{tabular}

\medskip
The averages are error-weighted and include all measurements,
including those taken during eclipses.  Relative errors for single
measurements are about 4.5\% in $JHK$ and 2\% in $U$ and 1\% in
$BVRI$.  The mid-IR errors are those of the average value. Upper
limits are 5$\sigma$.
\end{minipage}
\end{table*}

\subsection{Mid-IR photometry}
The mid-IR photometry was acquired with the Long Wavelength
Spectrometer (LWS, \citealt{jones}) which shares the forward
Cassegrain module of the Keck I with the Near InfraRed Camera (NIRC,
\citealt{matthews}). We used the $M$ (4.4-5.0~\um), SiC
(10.5-12.9~\um) and 17.6 filters (17.2-18.1~\um). Data were taken on a
128x128 Block Impurity Band Si:As infrared array (0.08\arcsec~ per
pixel) in chop-nod mode to correct for sky and telescope thermal
background. The frame integration time and number of coadds per
chop/nod were set to their nominal values by the controller software
once a filter and total integration time on object (excluding
overheads) had been chosen. We used combinations of several 100-400s
exposures to lower our detection threshold.

The average FWHM during the run was 4.3$\pm$0.9 pixels. We extracted
the count rate from a 16 pixel diameter circular aperture and
estimated any residual sky from a 20-30 pixel diameter annulus
centered on the source.  The count rates were converted to fluxes
using observations of $\alpha$ Ari for which an absolutely calibrated
spectrum\footnote{Electronic table available at {\sf
http://www.journals.uchicago.edu
/AJ/journal/issues/v117n4/980440/HD12929.tem}} has been established by
\citet{cohen}. The observations of $\alpha$ Ari were at airmass 1.0
and have a S/N $\sim 7600$ in $M$, 6400 in SiC and 540 with the 17.6
filter. Taking into account the filter transmission, the expected band
fluxes are 3.3~10$^{-15}$ \fw in $M$ band (4.8~\um), 1.29~10$^{-16}$
in SiC (11.7~\um) and 2.46~10$^{-17}$ in the 17.6~\um\ filter. The
absolute calibration error is 3.6\% ($M$) 2.5\% (SiC) and 3.4\%
(17.6). To correct for atmospheric extinction, we adopt the Mauna Kea
median values of \citet{krisciunas}, 0.22 mags/airmass at 4.8~\um,
0.15 at 11.7~\um\ and 0.45 at 17.6~\um\ (with a $\pm0.05$ mags/airmass
error on the mean). All the science exposures were taken at airmass
$\lta 1.3$. We checked our procedure by verifying that the derived
fluxes of the standards ($\gamma$ Dra, $\alpha$ Lyr, $\alpha$ Peg,
$\iota$ Cap, $\zeta^2$ Cet) observed at various times/airmass in
between science exposures were consistently found within 1$\sigma$ of
their expected values.

AE~Aqr is detected at 4.8 and 11.7~\um\ with a S/N$>$20. AE~Aqr
is also detected in three of the 17.6~\um\ exposures at a flux level
$>$ 40~mJy and a S/N$>$3 (once on Aug. 18 and twice on Sep. 16). The
source is easily identified in the corresponding sky-subtracted
images.  On Sep. 15, coadding the two best 17.6~\um\ exposures yields
a marginal 2.5$\sigma$ detection with a flux of 28$\pm 11$~mJy.

SS~Cyg is detected at 4.8 and 11.7~\um with S/N$>$10, but not detected
at 17.6~\um.  To increase sensitivity, we combined the four
consecutive 312~s exposures made on Sep.~15 and find 7.0$\pm$7.5~mJy
at the position of SS~Cyg. During those observations the sky steadily
increased by about 5\% as the airmass increased from 1.24 to
1.43. Combining the two 17.6~\um\ exposures obtained on Aug.~18 yields
6.7$\pm$15~mJy.

For a given LWS pointing, the source position on the array varies by
$\lta$ 1 pixel between filters. For AE~Aqr and SS~Cyg the source
location in the instrument field-of-view is known accurately from
several filters. For the non-detections of AE~Aqr or SS~Cyg in single
17.6~\um\ exposures, the count rate was integrated at the expected
location to determine an upper limit.  When the target is not detected
in any band, there is no {\em a priori} knowledge of the source
position on the detector (because of the wavelength regime and limited
field-of-view there is no other object in the image to use for this
purpose). In this case, we checked that there is no flux excess in the
image with an aperture photometry S/N$\gta$5 and, hence, adopted a
conservative 5$\sigma$ upper limit.

\subsection{The mid-IR variability of AE~Aqr and SS~Cyg}

\begin{figure}
\centerline{\epsfig{file=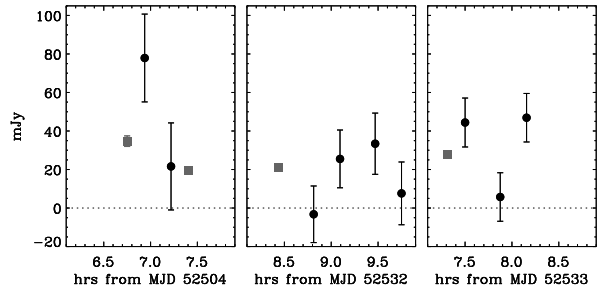,width=8cm}}
\caption{Variability of the AE Aqr mid-IR observations. Grey squares
are plotted for the four measurements at 11.7~\um\ (error bar smaller
than the square size), black circles for the nine 17.6~\um\
measurements. Note the variability of the 11.7~\um\ flux in the first
panel (Aug. 18) and 17.6~\um\ flux in the last panel
(Sep. 16). \label{fig:varaeaqr}}
\end{figure}

\begin{figure}
\centerline{\epsfig{file=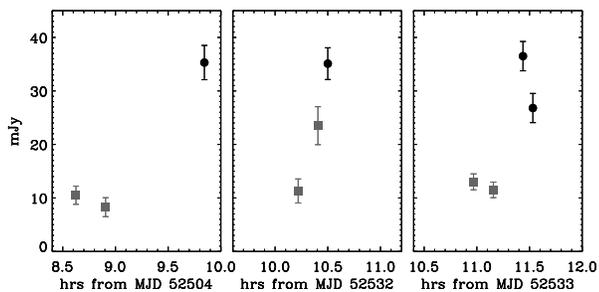,width=8cm}}
\caption{Variability of the SS~Cyg mid-IR observations. Grey squares
represent the six SiC (11.7~\um) measurements, black circles for the
four $M$ (4.8~\um) measurements. Note the 11.7~\um\ flux increase in
the middle panel (Sep. 15) and the 4.8~\um\ flux decrease in the last
panel (Sep. 16).
\label{varsscyg}}
\end{figure}

Both AE~Aqr and SS~Cyg show variations in the mid-IR fluxes measured
during the program. These are shown in
Figs.~\ref{fig:varaeaqr}-\ref{varsscyg}. Since sky transparency or
cirrus could be responsible, we inspected the variations in the
thermal background of the $M$, SiC and 17.6~\um\ filter chop-nodded
images. The night of Sep. 16 offered the best conditions for mid-IR
photometry: on Sep. 15 and Aug. 18 the mean sky level was about 15\%
higher (5\% in $M$). During the exposures, the thermal background was
stable to a level $\la 1$\% on Sep. 15-16 and 2-5\% on Aug. 18.

For AE~Aqr, the average sky level was constant during the three
consecutive 17.6~\um\ observations of Sep. 16 (airmass 1.1). The
non-detection during the second 17.6~\um\ exposure on Sep. 16 is
therefore very likely due to source variability. The variability at
4.8~\um\ ($\sim 30$\%, 4.5$\sigma$ away from the mean) and 11.7~\um\
($\sim$50\%, 5.5$\sigma$), where the source is well detected above the
background should not be affected by sky brightness changes.

The IR fluxes of SS~Cyg are constant but for two single measurements
which deviate by $\ga 3\sigma$ from the median flux: a factor 2
increase at 11.7~\um\ (3.2$\sigma$) and a 30\% decrease at 4.8~\um
(3.4$\sigma$), both on timescales of a few minutes. The measurements
are so close in time that the atmospheric extinction should not play a
role. For both observations the sky background was stable to better
than 1\% both within the exposure and compared to the
preceding/following exposure.

In both SS~Cyg and AE~Aqr there is a well-defined mean at 4.8 and
11.7\um. This strenghens our confidence in the data reduction / error
analysis and reinforces our conviction that the flux variations are
real, particularly when they occur within consecutive exposures on
short timescales.  Although there does not seem to be any hints of
varying sky conditions, a larger set of observations would place the
variability beyond any reasonable doubt.

\subsection{Near-IR photometry}
The $J$, $H$ and $K$ band photometry was obtained by switching to NIRC
\citep{matthews} either just before or after the LWS mid-IR
measurements. Data were taken on a 128x128 pixels (0.15\arcsec\ per
pixel) infrared InSb subarray with integration times of 0.038-1.5s to
avoid detector saturation. For each source, between 20 and 500 short
integrations were coadded to increase signal-to-noise at two telescope
nod positions 5\arcsec\ apart. Subsequent subtraction of the two
images removes the sky thermal contribution at the source position.

During the program, the average FWHM was 3.8$\pm1$ pixel. Count rates
were extracted from a 10 pixel diameter circular aperture.  Absolute
photometry was derived from observations of SJ 9107
($J$=11.934$\pm$0.005, $H$=11.610$\pm$0.004, $K$=11.492$\pm$0.011, see
\citealt{persson}).  Atmospheric extinction is corrected using the
median values observed at Mauna Kea (0.102 mag/airmass in $J$, 0.059
in $H$ and 0.088 in $K$, see \citealt{krisciunas}). We checked the
consistency of our calibration with four other standards observed
during the run. Comparing the count rates of the standard stars at the
two consecutive nod positions, we find the error budget is dominated
by a systematic uncertainty $\lta$ 0.05 mags. Individual $J$, $H$ and
$K$ band measurements for the cataclysmic variables are listed in
Tab.~1. Magnitudes are converted to fluxes in Jy using the California
Institute of Technology (CIT) zero points.

\subsection{Visible light photometry}
Optical images were taken in Johnson $U$, $B$, $V$ and Kron-Cousins
$R$, $I$ filters for each object at the Palomar 60-inch. Each object
was observed 1-4 times in each filter. Images were dark-subtracted,
flat-fielded using twilight exposures, and cosmic-ray
subtracted. Aperture photometry was performed with a 9.4\arcsec\
(25-pixel) radius aperture (except WZ Sge, described below),
estimating the sky background from an annulus with inner and outer
radii of 10\arcsec\ and 20\arcsec\ (27 and 54 pixels). The IDL
procedure aper, a component of the IDL Astronomy Users Library, was
used for the photometry and sky subtraction. \citet{landolt} standards
110 441, 115 420, and 110 502 (reduced using the same aperture
photometry as above) were used for absolute calibration and estimation
of color and extinction corrections.

The field of WZ~Sge was too crowded to use the same aperture, so a
2\arcsec\ aperture and sky annulus with radii of 10\arcsec\ and
5\arcsec\ were used.  Two nearby differential photometry standards,
stars 7 and 12 of \citet{misselt}, identical to stars 48 and 45 of
\citet{henden}, were analyzed using both the smaller and larger
apertures to estimate the aperture correction to match measurements of
the other objects and standard stars. A number of other differential
standard stars from \cite{misselt,henden}, in the fields of AE~Aqr,
SS~Cyg, RX~And, IP~Peg, and RW~Tri, were analyzed to ensure that
photometric conditions were consistently good throughout the observing
run.

\subsection{Visible and near-IR variability}
Some objects show small-amplitude variability in the visible, perhaps
due to flickering (usually interpreted as stochastic processes in the
accretion flow). This is most prominent at short wavelengths where
accretion dominates the SED. Four objects are also known to undergo
eclipses of the primary (HU~Aqr, IP~Peg, RW~Tri) or of the hot spot
(WZ~Sge). Using published ephemeris for these objects
\citep[respectively][]{ephhuaqr,ephippeg,ephrwtri,ephwzsge}, we find
that the $RIJHK$ data for HU~Aqr were taken at phases $0.9<\phi<1.0$,
the $UBVRI$ data for IP~Peg were taken at phases $0.8<\phi<1.1$ and
that the $I$ data for WZ~Sge is taken at phases $0.88<\phi<0.96$. All
of these measurements are likely to be affected by an eclipse.

\section{SECONDARY STAR CONTRIBUTION TO THE SPECTRAL ENERGY DISTRIBUTIONS}
\begin{figure*}
\centerline{\epsfig{file=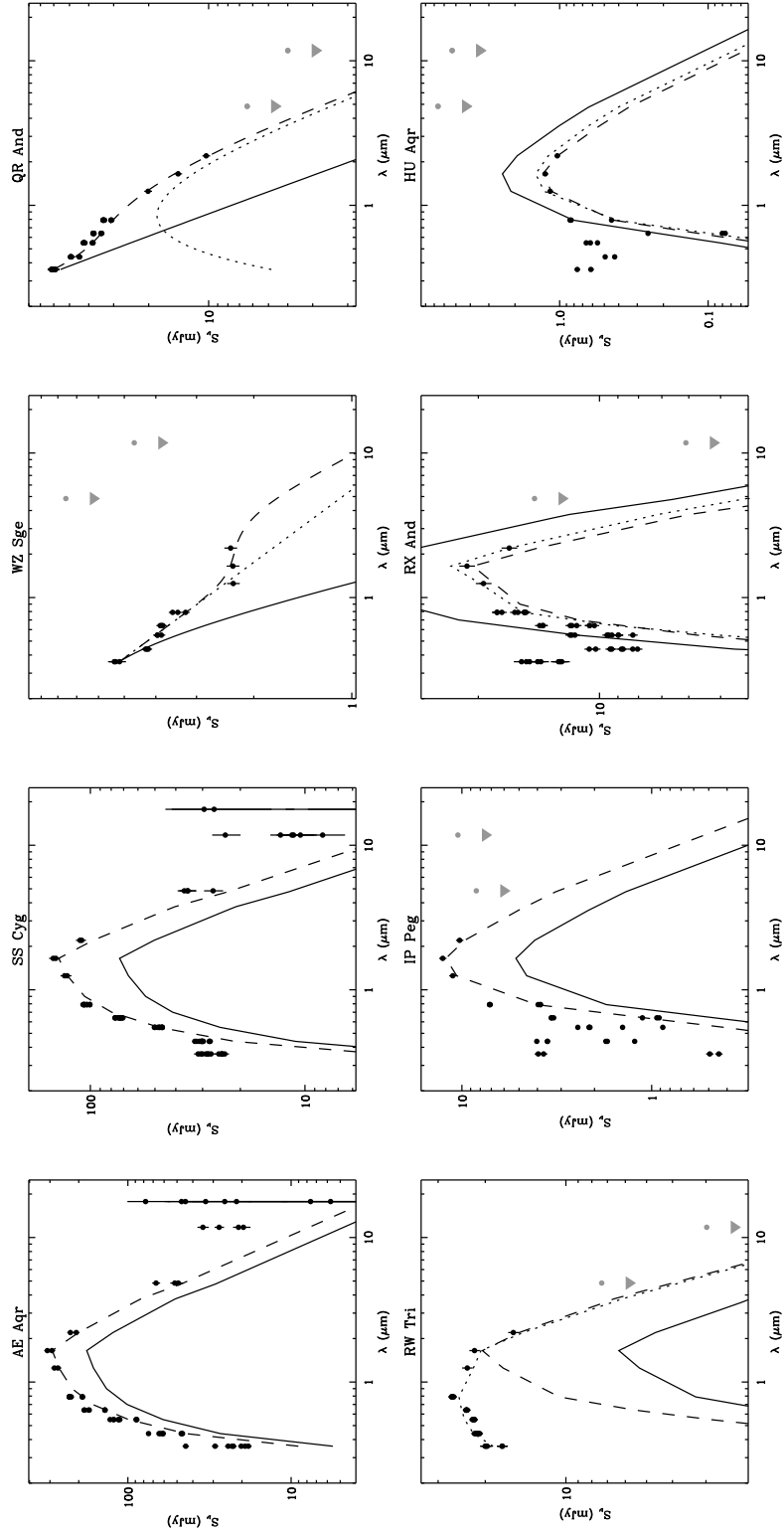,width=21cm,angle=90}}
\caption{Spectral Energy Distributions of the CVs observed during our
program. The bands covered are $UBVRIJHKM$, 11.7~\um\ and
17.6~\um. Every single measurement (and error bar) has been
plotted. Upper limits (arrows) are shown in grey. Solid lines
correspond to the expected SED of the secondary star under the
assumptions of Tab.~1; dashed/dotted lines correspond to SEDs with
different distance/spectral type (see \S3 for details). For RW~Tri,
the dotted line is a secondary + blackbody model (\S3.3). For WZ~Sge
(\S3.4) and QR~And (\S3.5) the solid line corresponds to the
Rayleigh-Jeans tail of the white dwarf SED; the dotted/dashed line
correspond to power law + blackbody models.\label{figsed}}
\end{figure*}

We constructed the spectral energy distribution of each system from
visible to IR wavelengths using this new data. All of our derived
fluxes, errors and upper limits are plotted in Fig.~\ref{figsed}. We
detect all of the objects in near-IR but only two objects at mid-IR
wavelengths, SS~Cyg and AE~Aqr (these measurements are discussed in
\S4 and \S5). In this section we analyze the visible and near-IR flux
densities in terms of the expected contributions from the secondary
star and investigate any long wavelength excess.

\subsection{Expected secondary star SEDs}
Our choice of plotting flux densities in mJy against wavelength in
Fig.~\ref{figsed} highlights the K/M secondary star contribution which
peaks in near-IR. Except for the short period WZ~Sge and the supersoft
QR~And, the donor star is easily identified as a bump in the SED. We
list in Tab.~1 the spectral type of the secondary in each system as
determined from previous visible/IR spectroscopy studies (references
are listed in Tab.~1). The uncertainty is typically one subtype.  In
principle, knowledge of the spectral type and distance allows one to
subtract the contribution of the secondary from the SED. To compute
expected secondary star SEDs we use the tables of colour and absolute
magnitudes as function of spectral type compiled in \citet[p.151 and
p.388]{astroq} with additional data from \citet{johnson} and
\citet{bessel}.  These tables cover the $U$ to $L$ (or $M$) bands. The
fluxes at longer wavelengths were extrapolated assuming a
Rayleigh-Jeans tail.

The expected SED of the secondary star under the assumptions of Tab.~1
is plotted as a solid line in Fig.~\ref{figsed}. For AE~Aqr, SS~Cyg,
IP~Peg and RW~Tri the expected flux is smaller than the observed
SED. For HU~Aqr and RX~And the expected flux is too large. In both
cases, the observed and expected near-IR SED can be matched by varying
the distance (dashed line in Fig.~\ref{figsed}). In so doing, we
assume the secondary contributes all of the near-IR flux from the
system hence {\em a priori} neglecting contributions from the
accretion flow, hot spot, white dwarf {\em and CB disc}.  The inferred
distance to the system is a lower limit. We obtain 80~pc for AE Aqr,
115~pc for SS~Cyg, 85~pc for IP~Peg, 190~pc for RW~Tri, 270~pc for
RX~And and 260~pc for HU~Aqr.

\subsection{IP~Peg, RX~And and HU~Aqr}
The distance listed in Tab.~1 for IP Peg is based on the mid-eclipse
$K$ band flux while our out-of-eclipse SED clearly includes flux from
accretion, thus making the source appear closer.  Assuming the
distance to IP Peg is 130~pc \citep{szkody}, we find the secondary
contributes $>$ 43\% of the near-IR flux while \citet{littlefair} find
that the secondary contributes about 62\% from $K$ band spectra. In
both RX~And and HU~Aqr, the expected secondary is too luminous at the
distance in Tab.~1. The distance to HU~Aqr is based on the $K$ band
flux and has an uncertainty compatible with our measurement
(191$^{+189}_{-115}$~pc). The distance to RX~And is based on the UV
spectrum of the white dwarf and can probably also accomodate the 35\%
discrepancy. Alternatively, a slightly later spectrum could reconcile
the observed SED and the distance in Tab.~1 for both objects (K7V
instead of K5V for RX~And and M4.5V instead of M4V for HU~Aqr, shown
as a dotted line in Fig.~\ref{figsed}). \citet{ciardi} find that the
eclipse $JHK$ fluxes of HU~Aqr are best fitted by a M6V star.

\subsection{RW~Tri, SS~Cyg and AE~Aqr\label{rwtri}}
The parallax distances for RW~Tri, SS~Cyg and AE~Aqr (respectively
341$_{-31}^{+38}$~pc, 159$^{+12}_{-11}$~pc and 102$^{+42}_{-23}$~pc)
allow a more accurate determination of the contribution from the
secondary. Our inferred distance is within the error on the parallax
distance of AE~Aqr, hence the secondary could well contribute all of
the near-IR flux from the system. For SS~Cyg and RW~Tri, our
photometric distance is much lower. Assuming the parallax distance and
spectral type of SS~Cyg and RW~Tri are correct\footnote{We note that
the parallax distance of SS~Cyg is inconsistent with the system being
a dwarf novae: within the usual assumptions of the disc instability
model, the luminosity would then imply a novae-like mass
transfer rate \citep{schreiber}.}, the secondary contributes 60\% of
the near-IR flux in SS~Cyg and 38\% in RW~Tri. Considering the
residuals this would leave for SS~Cyg, it is more likely that the
luminosity of the secondary is underestimated rather than there is a
large residual flux from accretion (in which case one would expect
power law residuals). By comparing such photometric estimates of the
fractional secondary contribution to others obtained by template
fitting of spectroscopic data, \citet{harrison} also find that the
secondary star of SS~Cyg is more luminous than its spectral type would
indicate. Secondaries in long-period CVs (such as AE~Aqr or SS~Cyg)
are likely to be evolved and have larger radii than main-sequence
stars of the same spectral type \citep{smith}.

For RW~Tri, a secondary contributing all of the $K$-band flux
would leave significant $UBVRI$ residuals. \citet{dhillon} find that
65$\pm 5$\% of the $K$-band light from RW~Tri is from the M0
secondary. Fixing the M0 contribution to 65\% of the $K$ band flux, we
find the residuals are adequately fitted by a blackbody of temperature
10,000~K and a peak flux at $\approx 0.5$~\um\ of 22~mJy. This
two-component model is shown by a dotted line in
Fig.~\ref{figsed}. The corresponding blackbody radius\footnote{Defined
so that the solid angle under which the blackbody is seen is $\pi
(R_{\rm bb}/d)^2$} $R_{\rm bb}\approx 2$~$10^{10} (d/341{\rm
~pc})$~cm is comparable to the expected size of the hot ($T_{\rm
eff}\ga 7000$~K) and stable accretion disc in this novae-like system
\citep{smak2}. RX~And and SS~Cyg which have comparable $P_{\rm orb}$
(comparable disc sizes) are dwarf novae observed in quiescence in
which the disc is cold.

\subsection{WZ~Sge}
There is no obvious signature of the secondary in WZ~Sge. The $U$ and
$B$ band fluxes of WZ~Sge agree well with a Rayleigh-Jeans tail (solid
line, Fig.~\ref{figsed}). The blackbody temperature would be $>$
15,000~K, in line with UV data fits which give a quiescent temperature
of $\approx$ 16,500~K for the white dwarf primary
\citep{smak}. However, there are other contributions at $V$ and longer
wavelengths which cannot be explained by a simple combination of the
Rayleigh-Jeans tail with another, lower temperature blackbody. Our SED
is identical to that of \citet{ciardi}, showing WZ~Sge has returned to
a similar quiescent state following its 2001 outburst. Following their
work, we also found that the overall SED could be reproduced by the
sum of a flattish power law ($S_\nu \sim \nu^{-0.6}$, dotted line in
Fig.~\ref{figsed}) and a blackbody with a temperature
$\la$1700~K. Previous studies have shown that most of the visible and
near-IR light in WZ~Sge could arise from the hot spot
\citep{spruit,ciardi}. The flat index of the power law may therefore
be linked to the hot spot rather than to the accretion disc. The
'flattening' at $JHK$ is due to the small contribution from the
secondary (dashed line in Fig.~\ref{figsed}). We find the blackbody
radius of the fitted 1700~K secondary is $R_{\rm
bb}\approx$0.1~R$_\odot$~($d$/48~pc), consistent with the expected
Roche lobe size \citep{smak}. The secondary would contribute $\sim$
20\% of the $K$-band flux, in rough agreement with spectroscopic
estimates that give $\la$ 10\% \citep{ciardi,dhillon}.

\subsection{QR And: signs of an F-subgiant secondary ? \label{qrand}}
In the supersoft source QR~And, the white dwarf has a temperature
$kT_{\rm wd}\ga 10$~eV. As in WZ~Sge, there are additional sources of
radiation above the Rayleigh-Jeans tail of the white dwarf
radiation. Unlike WZ~Sge, this extra component is very well modelled
by a single blackbody with a temperature of 6000~K. The corresponding
blackbody radius is $R_{\rm bb} \approx$3~R$_\odot$~($d$/2 {\rm
kpc}). This emission could arise from the accretion disc or the
secondary. A hot, optically thick disc (as expected in a supersoft
with a high mass transfer rate) would have a flatter $\nu^{1/3}$
spectrum blueward of the peak. On the other hand, the temperature and
radius correspond well to a late F subgiant filling its Roche lobe.

This is the type of secondary expected in models in which continuous
nuclear burning on the white dwarf is fueled by Roche lobe overflow
\citep{kahabka}. Light from accretion processes and from the white
dwarf swamps the stellar features from the companion (H and metal
lines) in visible spectra, rendering identification arduous. In our
spectral decomposition, the secondary contributes $\sim$ 80\% of the
light above 1~\um. The absence of ellipsoidal modulations would imply
a low system inclination. At the same time the system cannot be
face-on since eclipses of the hot inner disc are observed in UV
\citep{hutchings}.

We are aware of only one previous IR study of QR~And, that by
\citet{qf}. Their near-IR spectra show no lines from the secondary as
expected from a late F type. Their broad band SED does not show the
excess we identify. However, their photometry associates $JHK$ data
with $UBVRI$ data taken two years earlier and the non-simultaneity
could explain the discrepancy. Our data tentatively provides the first
hint of the secondary star in a supersoft source.

\section{MID-INFRARED EMISSION FROM AE Aqr}
\begin{figure}
\centerline{\epsfig{file=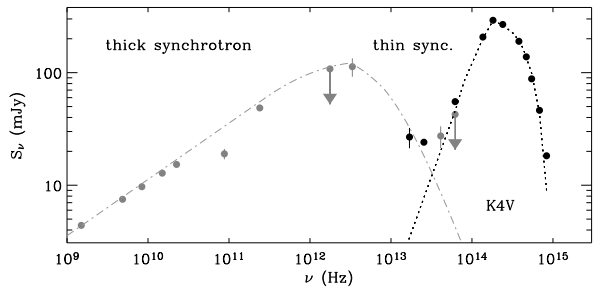,width=8.5cm}}
\caption{Average Spectral Energy Distribution of AE~Aqr. The visible
and IR fluxes from Tab.~2 are shown in black. The radio and ISO data
points shown in grey are from \citep{abada1,abada2}. Dotted line is
the K4V star SED normalised to fit the near-IR data. The low frequency
part of the SED from AE~Aqr has the telltale characteristics of
synchrotron emission. The dash-dotted line illustrates a 'synchrotron
clouds' model in which the low frequency part has an index 0.5 ($p=2$,
$\beta=0.6$ in Eq.~\ref{Eq:index}) and the high frequency part has an
index -5/3 corresponding to $p=2$ when synchrotron cooling
dominates.\label{fig:aqrsed} }
\end{figure}

The average mid-IR flux levels are well in excess of the
Rayleigh-Jeans tail from the secondary star, even when it contributes
all of the near-IR flux. At 4.8~\um, the lowest fluxes (50~mJy) are
compatible with the Rayleigh-Jeans tail but the highest flux (67~mJy)
is 4.5$\sigma$ above. At longer wavelengths the discrepancy is much
greater (see Fig.~\ref{fig:aqrsed}). The variability and 17.6~\um\
detection clearly require an additional source of radiation.

\subsection{Thermal circumbinary emission ?}
If the source is thermal, the increase in flux (upturn) from 11.7 to
17.6~\um\ suggests its temperature is in the range 200-300~K. Assuming
blackbody radiation, the required area would be $\sim
3~10^{25}$cm$^2$, larger than the binary size.

Based on the Sep. 16 observations, the timescale for the mid-IR
variability is $\la$ 15~mn.  It is doubtful that such a large emitting
area can coherently vary on such a short timescale. In a CB disc the
shortest timescale is the Keplerian timescale at the inner edge of the
disc and this is longer than the 9.88~hr system orbital
period. Flickering on timescales of mn with amplitude $\la 1$~mag in
the visible lightcurve of CVs has been attributed to the outer
accretion disc hotspot (where incoming material intersects the disc),
circumventing the Keplerian timescale lower limit. We cannot strictly
rule out similar flickering produced by material added in a
boundary layer at the CB disc inner edge or from MHD turbulence in the
CB disc. Reprocessing by circumbinary material of variable high energy
photons emitted deeper in the potential well is unlikely since the
mid-IR flux is of the same order as the UV/X-ray flux
($\sim$$10^{-11}$~erg~cm$^{-2}$ s$^{-1}$, \citealt{eracleous}). A 50\%
mid-IR variability would require changes of a factor 5 or more in the
irradiation luminosity. We conclude that the variability would be
quite outstanding if the flux is thermal. There is, however, a much
simpler non-thermal explanation.

\subsection{IR synchrotron from expanding clouds\label{sync}}
AE~Aqr is unique amongst cataclysmic variables for its radio
flares. These flares are interpreted as self-absorbed synchrotron
emission from expanding clouds of relativistic electrons propelled out
of the system by the fast-spinning white dwarf \citep{dulk}. These
same clouds can produce highly variable IR emission in their early
stages, which can explain our observations.

The time-averaged radio spectrum has a slope $\alpha\approx0.5$
\citep{abada1} flatter than the 2.5 slope expected from self-absorbed
synchrotron from a power-law distribution of relativistic
electrons. This likely results from the superposition of many flares
emitted at different times and cooling adiabatically. Within the
framework of \cite{vanderlaan} the spectral index $\alpha$ of the
summed contribution can be shown to have an asymptotic value of:
\begin{equation}
\alpha=\frac{5}{2}-\frac{(p+4)(1+3\beta)}{2\beta (2p+3)}
\label{Eq:index}
\end{equation}
where $p$ is the energy index of the electron power-law distribution
and $\beta$ characterizes the cloud expansion $R\propto t^\beta$
($\beta=1$ for steady expansion and $\beta=2/5$ for expansion in a
uniform medium). The 0.5 radio slope suggests decelerating blobs if,
as is likely, $p \ga 1$.

In Fig.~\ref{fig:aqrsed} we plot the SED including the radio and ISO
measurements (7.3, 90~\um) or upper limits (4.8, 170~\um) reported in
\cite{abada1,abada2}. Our observations fit in well, confirming the
synchrotron emission does extend to the IR in the optically-thin
regime. The variability and mid-IR fluxes are easily interpreted in
terms of this emission. If the synchrotron peak is at frequencies
close to 3300~GHz, then the initial electron cloud size is of the
order:
\begin{equation}
R\approx 10^9 ~B_{\rm 1000G}^{1/4} S_{\rm 0.1 Jy}^{1/2}\ \nu_{\rm 3300 GHz}^{-5/4} ~~{\rm cm}
\end{equation}
The adiabatic cooling timescale is of the order $\tau_{\rm ad}\sim
R/\beta v \sim$ 10~hr for an initial expansion speed of 1000~km/h.
The synchrotron cooling timescale is initially of order
\begin{equation}
\tau_{\rm syn}\sim 20~ B^{-3/2}_{1000G}\ \nu^{-1/2}_{\rm 3300GHz}~{\rm s}
\end{equation}
but increases quickly as the cloud expands ($B\propto
R^{-2}$). Synchrotron losses largely dominate in the IR, causing rapid
variability. In fact, the mid-IR probably varies on much shorter
timescales (seconds to minutes) than we observe.

With synchrotron losses dominating, the spectral slope in the
optically-thin regime steepens from $\alpha=-(p-1)/2$ to $-(2p+1)/3$
\citep{kardashev}. The IR spectral index can be estimated from the
consecutive 11.7~\um\ and 17.6~\um\ measurements taken at the
beginning of Aug. 18 and Sep. 16. We find $\alpha=-1.9\pm0.7$
(Aug. 18) and -1.1$\pm 0.7$ (Sep. 16) implying $p=1-3$, close to the
canonical $p=2$.

Our observations do not have the time-resolution and multiwavelength
simultaneity necessary to prove that the mid-IR emission does come
from ejected clouds. Such information could readily be obtained by the
{\em Multiband Imaging Photometer} flown on {\em SIRTF} and would
yield important information on the first stages of particle
acceleration/cooling. Brighter and closer than neutron stars in the
propeller regime, AE~Aqr is an ideal test-bed for studies of the
magnetic field - accretion flow interaction.

\section{MID-INFRARED EMISSION FROM SS~Cyg}
Inspection of the AAVSO lightcurve shows our observations of the dwarf
novae SS~Cyg occurred while the system was in quiescence.  Our SED is
consistent with that reported in quiescence up to the $L$ band
(3.4~\um) by \cite{szkody77}. We detect this system at 4.8~\um\ and
11.7~\um\ with fluxes of 33~mJy and 11.6~mJy respectively. Both fluxes
are well in excess of the mid-IR levels expected from extrapolating a
K4V secondary star normalised to the $H$ flux if we conservatively
assume that the secondary is contributing all of the near-IR
(Fig.~\ref{fig:sscygsed}). Even with accurate distances, deconvolving
the SED for the various contributions (accretion, white dwarf,
secondary...) introduces too many parameters to unambiguously identify
an excess at visible to near-IR wavelengths \citep[\eg\ ][]{harrison}.

Our maximum measured 11.7~\um\ flux (23.5$\pm$3.6~mJy) is 5$\sigma$
away from the expected secondary star Rayleigh-Jeans flux, clearly in
excess of our optimistic estimate of a few mJy. The measurements,
which were obtained in quiescence, are consistent with the upper
limits obtained by \citet{jameson} when SS~Cyg was not yet in full
outburst (see \S1). But the 11.7 and 17.6~\um\ measurements fall far
below the (conflicting) 12 and 25~\um\ {\em IRAS} detections in
outburst reported by \citet{jameson} and \citet{iras}. It remains
possible that the mid-IR flux increases considerably during outburst,
perhaps due to the IR tail of the much brighter accretion disc.

\subsection{Emission from circumbinary material ?}
The IR variability suggests another emission component is present
besides the secondary star. As SS~Cyg was in quiescence during the
observation, we consider it unlikely that the origin is the accretion
disc. Quiescent accretion discs have flat temperature distributions
(as inferred from eclipse mapping and theoretical models) and are
probably optically thin with $T< 3000$~K. A disc blackbody (see \S6)
with an outer radius equal to the white dwarf Roche lobe
(5~10$^{10}$~cm) and a uniform temperature of 3000~K still fails
to account for the {\em average} 11.7~\um\ flux by a factor $\ga 4$
(an order-of-magnitude with the highest flux).

If thermal, the extra component is more likely to peak at those
wavelengths where we see most of the variability, and hence is
characterized by a lower temperature.  A blackbody with a temperature
lower than 1000~K could account for the observations (see
Fig.~\ref{fig:sscygsed}). But with a constant emitting area $\ga
10^{23}$cm$^2$ (at 159~pc), the component is larger than the binary
size ($a\approx 1.5~10^{11}$cm) and thus not associated with the
accretion disc.

Just as we previously argued for AE~Aqr, such thermal circumbinary
emission is hard to reconcile with rapid flux changes. Naively, the
intrinsic emission from a CB disc is expected to be constant on
timescales shorter than the 6.6~hr orbital period. However, more
complex models taking into account MHD turbulence, mass input into a
fluctuating boundary layer or modes excited by tidal torques could
lead to rapid flickering in circumbinary material and cannot be ruled
out. A longer set of data would be needed to characterize this
variability (stochastic, periodic, intermittent, etc).

Reprocessing of the binary radiation seems difficult.  The IR
variability would imply large variations of the irradiating luminosity
by a factor $(F_{\rm max}/F_{\rm min})^4$ (the reprocessed flux
$F\propto T$ in the Rayleigh-Jeans regime). In quiescence, the high
energy photons from the boundary layer are emitted in X-rays with a
flux $\la 6$~$10^{-11}$~erg~cm$^{-2}$s$^{-1}$ \citep{wheatley},
which is only a factor of a few higher than the IR fluxes. We cannot
rule out that there is a large, varying EUV flux but this stretches
plausibility. Because of these arguments (and others in \S6), we
cannot make a solid case for thermal circumbinary emission and have
explored other possibilities.

\subsection{Free-free emission from a stellar wind ?\label{free}}
The white dwarf in SS~Cyg has a lower magnetic moment than the
intermediate polar AE~Aqr.  Since it is not detected in radio
\citep{cordova} and does not show the large flares that are so
prominent in AE~Aqr, the IR fluxes in SS~Cyg do not likely result from
matter ejected by a propeller.

An alternative explanation for the mid-IR excess involves
optically thick free-free radiation emitted by a wind from the
secondary star or accretion disc.  Magnetic braking of the binary
system by the secondary star wind is widely believed to be responsible
for the evolution of CVs above the period gap (see however, \S 1). The
free-free emission from a uniform, spherically symmetric ionised wind
leads to a power law spectrum in radio and IR with $S_\nu\propto
\nu^{0.6}$ \citep{panagia,wright}.

\begin{figure}
\centerline{\epsfig{file=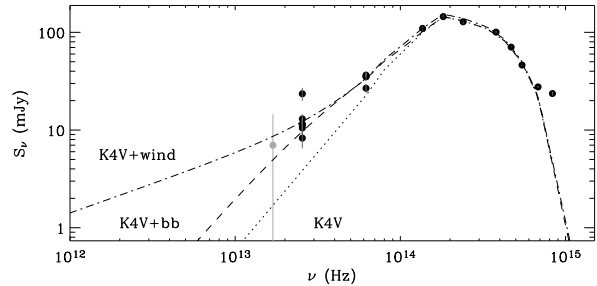,width=8.5cm}}
\caption{Average Spectral Energy Distribution of SS~Cyg. All of the
$M$ and SiC measurements have been plotted (only the lower envelope is
plotted for shorter wavelengths). The 17.6~\um\ flux is the limit
obtained by coadding the four Sep. 15 exposures. Dotted line is a K4V
star SED. The dashed line adds the contribution from a 1000~K
blackbody with an emitting area of $10^{23}$cm$^2$ (a single
temperature small CB disc). The dash-dotted line adds the free-free
emission from a wind with the parameters discussed in \S\ref{free}
(the wind emission is assumed to cut off at 2~\um).
\label{fig:sscygsed}}
\end{figure}

The median 4.8 and 11.7~\um\ fluxes of SS~Cyg give an index
$\alpha\approx 1.2$. However, the thermal spectrum of the secondary is
a significant contribution to the 4.8~\um\ flux. Assuming the median
10~mJy flux at 11.7~\um\ comes entirely from the wind, the expected
4.8~\um\ flux is $\approx$ 17~mJy. Since the median 4.8~\um\ flux is
35~mJy, this would imply the secondary star contributes about 18~mJy
at this wavelength, consistent with the Rayleigh-Jeans tail. At
17.6~\um, the expected flux would be $\approx 7$~mJy at the limit of
our sensitivity. The extrapolated flux would be 0.05~mJy at 6~cm,
which is consistent with the 0.1~mJy upper limit of
\citet{cordova}. The excess appears compatible with a $\nu^{0.6}$
spectrum extending to the radio. The proposed low frequency
contribution to the SED from wind free-free emission is shown in
Fig.~\ref{fig:sscygsed}

The wind mass loss rate required to explain the IR flux is \citep{wright} :
\begin{equation}
\dot{M}_{\rm wind}\approx 2~10^{-9} ~v_{\rm 50 km/s} S_{\rm 10mJy}^{0.75} \lambda_{\rm 11.7 \mu{\rm m}}^{-1/2} ~~{\rm M}_\odot~{\rm yr}^{-1}
\end{equation}
where we have assumed a wind speed of 50~km~s$^{-1}$ and a wind
temperature of 10,000~K. The corresponding radius of the emitting
region at 11.7~\um\ is about $4~10^{10}$~cm which also happens to be
the size of the Roche lobe of the star and comparable to the outer
radius of the accretion disc.  Variations in the IR flux could well be
attributed to inhomogeneities in the temperature, density or
ionisation of the wind close to the launching region. Alternatively,
the variations could be due to coherent emission during flares. This
would probably require very powerful magnetic fields (10$^7$~G, white
dwarf origin ?) for the gyrofrequency to be in IR.

The derived wind mass loss rate is very high, orders of magnitude more
than what would be expected from a field red dwarf and of the same
order as the mass loss through Roche lobe overflow. UV resonance lines
from a disc wind are observed during outbursts of dwarf novae but
these have lower mass loss rates and higher speeds by several orders of
magnitudes and we can only speculate as to the connection with the
present observations. The assumptions underlying our estimate may be
too simplistic. The IR photons are emitted close to the launching
region, not only where the wind is still accelerating but also where
spherical symmetry breaks down. The rapid rotation of the phase-locked
secondary and/or the complex configuration of the magnetic fields will
give substantial deviations from the time-averaged, simple model used
above. A fast magnetic rotator in this context could give rise to
flares of enhanced wind mass loss rate compared to the Sun. Our
estimate should therefore be considered with caution. Observations at
longer wavelengths, which probe a larger volume and will be less prone
to the caveats described, are urged to establish the nature of the IR
excess and (eventually) to give a better estimate of any wind mass
loss.

\section{LIMITS ON CIRCUMBINARY MATERIAL}

CB discs, once in place around a binary, have a nearly indefinite life
time and there are several avenues by which material could end up in a
CB disc during the birth and evolution of a binary (see \S1). CB discs
are a tantalizing possibility for explaining several anomalies in CV
phenomenology, but direct observation is the only way to firmly
establish their presence or absence. In this section we explore in
some detail what we can deduce in this respect from the observations
presented above. The conclusions of this section are compiled in
\S6.5.

We start by showing that if there is a significant amount of
circumbinary material then it has to be in a circumbinary (CB)
disc. By using the expected SED, the observations can limit the
maximum temperature in the disc. At the same time, passive
reprocessing of the binary light gives a minimum temperature of the CB
disc.  Within the uncertainties in the distance and the inclination
angle of the systems, our results cannot definitively indicate the
presence or absence of a CB disc.  The CB discs in these systems may
be shadowed, or have their inner regions vaporized and devoid of dust.
Any sizeable CB disc in the systems we observed is likely to be
predominantly cold and/or optically thin with a peak thermal emission
at longer wavelengths than we observed. We conclude on the
observational prospects for CB discs.

\subsection{IR emission from optically thin material and limits on dust mass\label{thin}}
Cataclysmic variables do not show large visual extinctions, even at
high inclinations. In our sample, the most reddened object is the
eclipsing source RW~Tri, which only has $E(B-V)=0.1\pm0.05$
\citep{verbunt}. Any circumbinary material covering a large solid
angle must therefore be optically thin \citep[this is briefly
mentioned by][]{bsc}. At low temperatures dust dominates the opacity
so the requirements that $\tau\ll 1$ translates into an upper limit on
the column number density $N$ of grains of size $a$:
\begin{equation}
\label{tau}
\tau_d=N\pi a^2 \bar{Q}(a,T_\star)\ll 1
\end{equation}
where $\bar{Q}(a,T)\approx {\rm min}\{1,0.072 ~a~T^{1.65}\}$ ($a$ in
cm, $T$ in K) is the Planck-averaged absorption efficiency of grains
with respect to radiation of temperature $T$. We use this upper limit
below.

Dust will be destroyed if its temperature is above the sublimation
point $\sim 2000$~K. The radiative equilibrium of dust grains at a
radius $R$ heated by radiation of temperature $T_\star$ and luminosity
$L_\star$ can be written as \cite[\eg][]{clayton}:
\begin{equation}
\pi a^2 \frac{L_\star}{4\pi R^2} \bar{Q}(a,T_\star) =4\pi a^2 \sigma T_d^4 \bar{Q}(a,T_d)
\label{eq:dust}
\end{equation}
where $a$ is the size of the grains and $T_d$ their temperature. The
dust temperature for $L_\star$=10$^{32}$~erg~s$^{-1}$ at 1~AU will be:
\begin{equation}
\label{tdust}
T_d \la 590~L_{\star, 32}^{0.18} ~R_{\rm AU}^{-0.35}~~{\rm K}
\end{equation}
where the upper limit comes from taking $\bar{Q}=1$ for absorption
(high $T_\star$) and $a=0.005$~\um, characteristic of condensation
nuclei. Larger $a$ or lower $\bar{Q}$ yield lower $T_d$. The
integrated optical to near-IR luminosities we observe range between
10$^{31-32}$~erg~s$^{-1}$ except in QR~And ($\sim
10^{35}$~erg~s$^{-1}$). The luminosity from accretion in a cataclysmic
variable (emitted in UV) reaches at most a few
$10^{34}$~erg~s$^{-1}$. Nova episodes during which the luminosity
increases to Eddington values may periodically destroy the grains at
radii $\la 10$~AU. In QR~And nuclear burning is continuous so that the
UV flux is probably too high for dust to be present close to the
source.  Nova outbursts occur on timescales of 1000 - 10000~years,
which should leave time for dust to reform in the circumbinary
material. For a typical CV, we conclude the temperatures are likely
sufficiently below sublimation (Eq.~\ref{tdust}) for dust to exist at
radii $\ga 5~10^{11}$~cm.

The settling timescale for a condensation nucleus ($a=0.005$~\um,
$\rho=2.25$~g~cm$^{-3}$) is 
\begin{equation}
\tau_{\rm settling}\sim \Sigma/a\rho \Omega_K\sim 8000~{\rm years}
\end{equation}
at the inner edge of a CB disc (10$^{11}$~cm) with
$\Sigma=1$~g~cm$^{-2}$. This is usually taken to give an idea of the
timescale on which larger grains might grow. However, there is
considerable uncertainty since turbulence, radial drift, drag,
coagulation and others are likely to change how large the dust can get
and on what timescale. Whether dust can form at all in a CB disc is an
open question. Note that the material from which the CB disc is formed
could be intrinsically dusty if it comes from the remnants of the
common envelope phase. Alternatively, observations of some novae show
that $\sim 10^{-5}$~M$_\odot$ of dust is formed in the shell ejecta
within days of outburst. Most of the ejecta has escape velocity but a
small fraction at the low velocity end could stay bound to the system
and provide a way to fuel the CB disc in dust. In the following, we
assume dust is present when the temperatures are below sublimation.

The flux density $S_\nu$ at a frequency $\nu$ due to optically thin
dust at temperature $T_d$ and distance $d$ to observer is:
\begin{equation}
S_\nu= \pi (a/d)^2 N_d Q_\nu(a,T_d) B_\nu(T_d)
\end{equation}
where $(\lambda/a)Q_\nu \approx 1-10$ at 10~\um\ \citep{draine} and
$N_d$ is total number of grains. At 11.7~\um, $S_\nu$ is maximized at
$T_d\approx 200$~K. The characteristic radius $R_d(T_d)$ of the
dust is given by Eq.~\ref{eq:dust}. Since $N_d\approx \pi R_d^2 N$
then, using Eq.~\ref{tau}, we rewrite the flux density as a function
of $T_d$:
\begin{equation}
\label{snu}
S_\nu=\tau_d \frac{L_\star}{16\sigma T_d^4 d^2} \frac{Q_\nu(T_d)}{\bar{Q}(T_d)} B_\nu(T_d)
\end{equation}
and the requirement that $\tau_d<1$ gives an upper limit on
$S_\nu$. Note that the dependence on $a$ cancels out so that this
depends only on $d$, $L_\star$ and $T_d$. Our observational upper
limits at 11.7~\um\ are about 7~mJy. Depending upon whichever is lower
for the parameters considered, we take this observed value or that
given by Eq.~\ref{snu} at 11.7~\um\ (with $\tau_d$=0.1) and calculate
the maximum allowed dust mass $M_d=4/3\pi a^3 \rho N_d$. Fixing
$d=100$~pc and the grain density $\rho \approx$ 2.25~g~cm$^{-3}$, we
find that the maximum dust mass at radii less than 10~AU is very small
$\la 10^{-9}$~M$_\odot$ for $T_\star\ga 2000$~K and $L_\star\ga
10^{32}$~erg~s$^{-1}$, decreasing rapidly with $T_\star$.

Using (1) that the extinction to CVs is low and (2) our IR upper
limits, we find there cannot be a significant amount of optically thin
material within 10~AU covering a large solid angle as viewed from the
source. If a large mass of bound material has piled up at circumbinary
radii then it must have formed an optically thick, geometrically thin
disc. Even then, the low extinction hints that the circumbinary discs
cannot be too massive: for instance, \citet{bertout} finds that on
average more than 20\% of T~Tauri circumstellar discs irradiated by a
K7V star should be extinguished if their mass flow rates $\dot{M}\ga
10^{-8}$~M$_\odot$~yr$^{-1}$. At most, CVs show only a weak trend of
being redder with higher inclination \citep{2mass}.

The small amount of matter surrounding CVs also gives some ideas on
mass transfer onto this circumbinary disc. An upper limit on the mass
transfer rate at a given radius $R$ is set by dividing $M_d(R)$ by the
Keplerian orbital timescale at $R$. Most of the matter leaving the
binary is added at small radii where the orbital timescale is
short. $\dot{M}_d$ is $\la 10^{-17}$~M$_\odot$~yr$^{-1}$ at radii
$R\la 0.1$~AU if the dust reprocesses efficiently a
10$^{32}$~erg~s$^{-1}$ luminosity.  Assuming a dust-to-gas ratio of
1\% (typical of protoplanetary discs) the circumbinary mass transfer
rate is very small, probably much less than 0.1\% of the mass transfer
rate between the secondary and white dwarf.  We note, however, that
these limits do not apply if (i) the dust-to-gas ratio in the
transfered material is very small because dust is vaporised or has not
had sufficient time to condense and/or (ii) the transfer occurs within
a small opening angle within the orbital plane.

\subsection{SED of optically thick CB disc\label{sed}}
The transport of angular momentum in a geometrically thin, optically
thick circumbinary disc generates `viscous' heat which is radiated
thermally. Theoretical models show that the optically thick region of
a CB disc expands to AU scales on timescales of tens of Myrs
\citep{dubus,taamsd}. The luminosity and SED of the disc will depend
on its (uncertain) evolutionary stage but the effective temperature
$T_{\rm eff}$ can be written as a power law:
\begin{equation}
T_{\rm eff}(R)=T_{\rm in}(R_{\rm in}/R)^{n}\label{Eq:teff}
\end{equation}
with $R_{\rm in}$ the inner radius of the circumbinary disc, $T_{\rm
in}=T_{\rm eff}(R_{\rm in})$ the effective temperature at this radius
and $n$ the index. Models show $T_{\rm eff}$ is initially steep with
$n\approx 2$. As the disc evolves, the mass flow rate becomes almost
constant in the optically thick region, approaching the steady-state
Shakura-Sunyaev index of 3/4 \citep{dubus}. Assuming each annulus in
the optically thick disc radiates as a blackbody, the expected SED is
the standard:
\begin{equation}
S_\nu=\frac{4\pi h \nu^3 \cos i}{c^2 d^2}\int_{R_{\rm in}}^{R_{\rm out}}\frac{R {\rm d}R}{{\rm e}^{h\nu/kT(R)}-1}
\label{limit}
\end{equation}
where $i$ is the binary inclination, $d$ the distance and $T(R)$ is
given by Eq.~\ref{Eq:teff}. For intermediate frequencies $kT(R_{\rm
out}) \ll h\nu\ll kT(R_{\rm in})$ the spectrum is a power law
$\nu^{3-2/n}$. At smaller frequencies the spectrum has a
Rayleigh-Jeans tail $\nu^2$ and at higher frequencies the spectrum has
a Wien cutoff. As $n$ increases or when the disc is not very extended,
the spectral window where the $3-2/n$ power law approximation is valid
is reduced and the SED approaches a blackbody.

\subsection{Upper limit on the CB disc temperature from mid-IR fluxes\label{sedul}}
The CB disc inner radius is set by tidal truncation at $R_{\rm
in}\approx 1.7 A$ where $A$ is the binary separation and the origin is
at the binary centre-of-mass \citep{artlub,gk}.  $R_{\rm in}\sim
10^{11}$~cm for the systems we surveyed (the values are given in
Tab.~\ref{tab:cbt}). At a given frequency $\nu$ and for a given index
$n$, the expected flux as given by Eq.~\ref{limit} depends only on
$T_{\rm in}$ and $R_{\rm out}$. $R_{\rm out}$ determines only the
location of the Rayleigh-Jeans tail at low frequencies and our
observations are insensitive to it. Therefore, for a given $n$, the
observed SEDs allow us to place an upper limit on $T_{\rm in}$.

\begin{figure}
\centerline{\epsfig{file=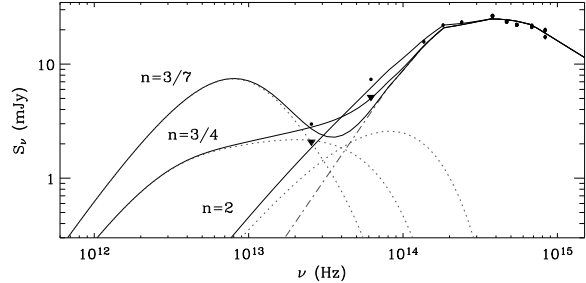,width=8.5cm}}
\caption{Circumbinary disc contribution to the SED allowed by the
observations of RW Tri. The dash-dotted line is the secondary star +
10,000~K blackbody model discussed in \S\ref{rwtri}. The CB disc is
tidally truncated at 1.9$~10^{11}$cm and the inclination is
70$\degr$. Various possible CB disc SEDs (\S\ref{sed}) are shown as
dotted lines. Their sum with the two-component model is plotted as a
solid line. For $T\propto R^{-3/4}$ ($n$=3/4, appropriate for a large
CB disc) the maximum allowed temperature at $R_{\rm in}$ is
1070~K. For $n$=2 (small CB disc) $T(R_{\rm in})$ is 2150~K. For
$n$=3/7 (flared reprocessing disc, see \S\ref{reproc}) $T(R_{\rm in})$
is 610~K. The outer radius of the CB disc, which determines the
location of the Rayleigh-Jeans tail at low $\nu$, is arbitrarily set
at 1~AU.\label{figrw}}
\end{figure}

As an example we take RW~Tri and plot in Fig.~\ref{figrw} the expected
SEDs obtained by finding the maximum $T_{\rm in}$ consistent with the
observed mid-IR fluxes, for various values of $n$. For RW~Tri we have
$R_{\rm in}\approx 1.9~10^{11}$~cm, $i\approx 70\degr$ and we
arbitrarily choose $R_{\rm out}=1$ AU (a higher $R_{\rm out}$ would
move the Rayleigh-Jeans tail to lower $\nu$ in Fig.~\ref{figrw}).  The
maximum allowable $T_{\rm in}$ is 2150~K for $n$=2, 1070~K for $n$=3/4
and 610~K for $n$=3/7. A steep temperature distribution decreases the
sum of the blackbody contribution at a given frequency and allows
higher temperatures. Except for high $n$, the observations require the
disc SED to be in the Wien regime at mid-IR wavelengths: our upper
limits `force' the peak contribution from the CB disc to longer
wavelengths. The visible and near-IR flux would be grossly
overestimated if the temperature were high enough to have the mid-IR in
the $3-2/n$ power-law regime of the disc SED.

Tab.~\ref{tab:cbt} lists the maximum temperatures that can fit the
observations of the various systems if we assume a disc with $n$=3/7
(an optically thick irradiation dominated isothermal CB disc, see
\S\ref{reproc}) with $n$=3/4 (appropriate for a large, evolved viscous
CB disc) or $n$=2 (representative of a small, unevolved CB disc). The
distance and inclination angle have the values listed in
Tab.~\ref{tab:cbt}.  We omit AE~Aqr as the evidence clearly points
towards a different origin for the mid-IR emission. In all the
systems, the CB discs must have low temperatures in order to fit the
observations. Comparison with theoretical models suggests that,
because of the temperature limit, the putative CB discs in our systems
would not be large and massive.

\subsection{Lower limit on CB disc temperature from stellar irradiation\label{reproc}}
Protoplanetary discs are very similar in size, density and temperature
to large, evolved circumbinary discs.  The visible to near-IR SEDs of
CVs and T~Tauris are comparable: the distance to the objects ($\sim
100$~pc) and spectral type of the stellar component (K-M) are not very
different. T~Tauris show large IR excesses associated with thermal
emission from circumstellar dust heated by stellar radiation
\citep[see \eg][]{rucinski}. Reprocessing can largely enhance the
emission above that expected from `viscous' heating alone.

The temperature in a flat disc reprocessing light from a star of
radius $R_\star$ and temperature $T_\star$ is \citep{friedjung2,shu}:
\begin{equation}
T_{\rm irr}^4=\frac{T_{\rm \star}^4}{\pi}\left[\arcsin r-r (1-r^2)^{1/2}\right]
\label{tirr}
\end{equation}
where $r=R_\star/R$.  At large radii $T_{\rm irr}\propto R^{-3/4}$
mimicking a viscous disc.  Any other source of heat (viscosity) will
raise the temperature above this minimum. Such a disc can reprocess up
to 25\% of the stellar radiation. The assumption of a flat disc is
acceptable if most of the dust (dominating the opacity) has settled
down to the midplane. However, irradiation heating will likely puff up
the disc and enable more of the stellar flux to be intercepted. Dust
in the optically thin upper layers of the disc will also raise the IR
flux \citep{chiang}. Taking the flaring into account, a steady,
optically thick, passive disc dominated by the reprocessing of light
from a point source will have a height $H\propto R^{9/7}$ and a
temperature distribution $T\propto R^{-3/7}$, much shallower than that
of Eq.~\ref{tirr}. 

Neglecting flux from the primary and other sources, the minimum
irradiation is that from the secondary. The time-averaged flux is
approximately equivalent to placing the secondary star at the binary
centre-of-mass and irradiating the inner edge of the CB disc at a
distance of 1.7$A$.  If all of the stellar radiation is absorbed by
dust and the CB disc is not shadowed (we come back on these
assumptions below), the minimum $T_{\rm in}$ will be given by $T_{\rm
irr}(R_{\rm in})$ (Eq.~\ref{tirr}). We take the Roche lobe radius for
$R_\star$ and the effective temperature corresponding to the secondary
spectral type for $T_\star$. The value of $T_{\rm irr}$ for each
system can be found in Tab.~\ref{tab:cbt}.

Fits to the observed SEDs with a multicolour disc (Eq.~\ref{limit})
give upper limits to $T_{\rm in}$ for given power-law distributions
which must be compatible with the expected minimum temperature from
irradiation $T_{\rm irr}$. The results of Tab.~\ref{tab:cbt} shows
that the lower limits, $T_{\rm irr}$, are inconsistent with the upper
limits, $T_{\rm in}$, for irradiation-dominated discs (those with
$n=3/7$), in all cases except HU~Aqr and WZ~Sge, the systems with the
shortest $P_{\rm orb}$. Consistency requires steeper values of $n$,
hence some intrinsic viscous heating. In most systems values of $n$
between 3/7 and 3/4 (the temperature distribution of a stationnary
optically thick Shakura-Sunyaev disc) are sufficient. In SS~Cyg and
RX~And steeper temperature distributions are required, implying small
CB discs. Any flaring will increase the value of $T_{\rm irr}$,
although not above that given by geometric dilution $T_\star
(R_\star/R_{\rm in})^{0.5}$ which are a factor $\sim$ 2 higher than
the values in Tab.~\ref{tab:cbt}. Additional contributions to
irradiation from the white dwarf or accretion flow will also increase
this minimum value. Higher values of $T_{\rm irr}$ will require even
higher values of $n$ for consistency in systems other than SS~Cyg and
RX~And.

\begin{table}
\begin{center}
\caption{Constraints on the CB disc temperature.}
\label{tab:cbt}
\begin{tabular}{lcccrrrr}
\hline
 Object & $R_{\rm in}$ & $R_{\star}$ & $T_\star$ & $T_{\rm irr}$ & \multicolumn{3}{c}{max $T_{\rm in}$}\\
 & \multicolumn{2}{c}{\small (10$^{10}$cm)} &  \multicolumn{2}{c}{\small (K)} & \multicolumn{3}{c}{\small $n$=\{3/7,3/4,2\}}\\
\hline
SS Cyg & 26.1 & 5.1 & 4340 &  870 & 450 &  630 & 1300 \\
WZ Sge & 6.8  & 0.8 & 1500 &  220 & 480 &  760 & 1000 \\
QR And & 47.6 & 12. & 6000 & 1490 & 900 & 1900 & 2700 \\
RW Tri & 18.8 & 4.5 & 3800 &  890 & 610 & 1070 & 2150 \\
IP Peg & 17.7 & 3.5 & 3180 &  640 & 550 &  930 & 1300 \\
RX And & 20.6 & 3.7 & 4340 &  820 & 500 &  770 & 1500 \\
HU Aqr & 10.1 & 1.5 & 3180 &  520 & 840 & 1100 & 1250 \\
\hline
\end{tabular}
\end{center}

\medskip
$R_{\rm in}$ is the inner CB disc radius. $R_{\star}$ is the secondary
star Roche lobe radius. $T_\star$ is the effective temperature
corresponding to the secondary star spectral type
(Tab.~\ref{tab:obj}). $T_{\rm irr}$ is the expected temperature at
$R_{\rm in}$ for a flat disc reprocessing the stellar radiation, which
represents a temperature lower limit (\S\ref{reproc}). $T_{\rm in}$ is
an upper limit to the temperature at $R_{\rm in}$ allowed by our
observations for a CB disc with a temperature distribution $T\propto
R^{-n}$ (\S\ref{sedul}). The last three columns are for (left to
right) $n=3/7$, $n=3/4$ and $n=2$. $R_{\rm out}$ was set at 1 AU. The
binary parameters used (mass, $q=M_2/M_1$, $i$, $P_{\rm orb}$) are
taken from \citet{ritter}. For QR~And we took $M$=2M$_\odot$, $q$=2,
$i$=60$\degr$ and $T_\star$=6000~K (\S\ref{qrand}).
\end{table}

Our analysis regarding reprocessing may overestimate its effect if
most of the CB disc is shadowed and is not exposed to the light
emanating from the binary system (secondary or other
sources). Provided that the shadowed region is cold enough that its
intrinsic emission has a minor impact in the IR, the overall effect
would be identical to having a very small $R_{\rm out}$. The maximum
allowable $T_{\rm in}$ is insensitive to $R_{\rm out}$ except when the
disc becomes small, and the small area allows high temperatures while
keeping the fluxes low. The pressure scale height variation in CB disc
models does suggest that the outer disc is shadowed
\citep{dubus}. Detailed radiative transfer calculations are needed to
verify that the inner edge of the CB disc can indeed shadow the rest
of the material.

An alternative to shadowing would be a dust-free disc. As discussed in
\S6.1, it is unclear whether dust in CB discs can form in situ. Since
the absorption is essentially due to grains, the efficiency with which
radiation is reprocessed (or, equivalently, the inner radius $R_{\rm
in}$ of the optically thick region) would change if the inner CB disc
was dust-free. The situation would be comparable to the circumstellar
discs of Herbig Ae/Be stars where the large radiation vaporizes the
dust at small radii. An optically thin layer of grains can be
vaporized up to $\sim 5~10^{11}$~cm for an average irradiating
luminosity of $10^{32}$~erg~s$^{-1}$ (see \S\ref{thin}). If there is
no reprocessing up to this radius then agreement between $T_{\rm in}$
and $T_{\rm irr}$ at the optically thick disc inner edge is easier to
achieve. A larger $R_{\rm in}$ requires a lower temperature to fit the
IR but the upper limits are looser as the peak thermal emission moves
to the far-IR, leaving more room for adjustment between the (optically
thick) lower limit $T_{\rm irr}$ and upper limit $T_{\rm in}$.

\subsection{Summary: prospects for detecting CB discs}
We summarize our results before addressing the prospect for finding CB
discs.
\begin{enumerate}
\item Distributions for circumbinary material other than a thin disc
lead to high extinctions or negligible mass.
\item The observations limit the maximum $T_{\rm eff}$ in an optically
thick circumbinary disc.
\item The maximum temperature is approximately consistent with the
minimum expectations from passive reprocessing of the binary
emission. In SS~Cyg and RX~And the temperature distribution must
be steep, implying small CB discs. This will generalise to other
systems if we underestimated reprocessing.
\item A small IR luminosity from the CB disc may result from a small
or shadowed disc or one in which its inner regions are vaporized.
\end{enumerate}

The contribution from any sizeable CB disc will be significantly
affected if it is either shadowed or has little dust in the
IR-emitting inner region. In these cases most of the matter will be
cold or optically thin, resulting in a shift of the peak emission
to longer wavelengths.  Observations in far-IR or sub-mm wavelengths
would then be more appropriate.

Narrow lines centered at the systemic velocity in high inclination
systems may prove a different, more amenable test of circumbinary
discs than their continuum emission (\S1).  For example,
\citet{mouchet} reported H$_2$ narrow absorption lines in the {\em
FUSE} UV spectrum of the polar BY~Cam that are intrinsic rather than
interstellar.  This is particularly important since molecular hydrogen
is a clear signature of cold, dusty material. The total column density
of H$_2$ in BY~Cam is high: $N_{\rm H_2}\approx
3~10^{19}$~cm$^{-2}$. \citet{hutchings} also report H$_2$ absorption
in their {\em FUSE} spectrum of QR~And with $N_{\rm H_2}\approx
10^{20}$~cm$^{-2}$. However, it is not clear that the absorption is
intrinsic in this system.  The more accessible H$_2$ 1-0 S(1) lines at
2.122~\um\ could be more fruitful than space-based UV in a systematic
search for molecular hydrogen in other CVs.

The molecular hydrogen column density can be used to estimate (for
instance) the $^{13}$CO J=1-0 line at 110.2~GHz in a reverse fashion
from what is usually done to interpret sub-mm observations of
circumstellar discs \citep[\eg][]{scoville,sargent}. The expected
integrated line flux is:
\begin{equation}
\int S_\nu dv \approx 5~\frac{{N}_{\rm H_2} [^{13}{\rm CO}]}{\theta_{\rm S}^2} \frac{e^{-5.29/T_{\rm x}}}{T_{\rm x}+0.88}\frac{1-e^{\tau}}{\tau}  ~{\rm mJy~km~s}^{-1}
\end{equation}
where $N_{\rm H_2}$ is measured in units of 10$^{19}$~cm$^{-2}$,
$[^{13}{\rm CO}]$ is the abundance of CO compared to H$_2$ (taken to
be 10$^{-4}$), $T_{\rm x}$ is the excitation temperature of CO, $\tau$
its optical depth and $\theta_{\rm S}$ is the angular size of the
emitting source.  Taking $\theta_{\rm S}$=1\arcsec\, $T_{\rm x}$=30~K
and $\tau\ll 1$ to estimate the line flux, $^{13}$CO emission could be
detected in BY~Cam ($\sim$15~mJy~km~s$^{-1}$) with present
instrumentation.

\section{CONCLUSIONS}
A search for mid-IR excess emission in several cataclysmic variables
has been undertaken, motivated by the possibility that some of them
may contain large circumbinary discs influencing their evolution.
Simultaneous visible to mid-IR photometry were obtained from which we
constructed spectral energy distributions.

We find that the near-IR SEDs of the observed CVs are reproduced by a
stellar SED with the spectral type expected from the secondary. Given
the uncertainties on distance and that secondaries in CVs may be more
luminous than their isolated main-sequence counterparts, we conclude
that there is no need for extra components in the near-IR. In RW~Tri,
a 10,000~K blackbody is needed in addition to the M0 secondary to
explain the large visible flux. In WZ~Sge, the SED can be reproduced
by a power law (accretion or hot spot) + a cool blackbody for the
faint, low-mass secondary. In QR~And, the SED is decomposed into a
Rayleigh-Jeans tail from the hot white dwarf and a blackbody of
temperature 6,000~K and radius 3~R$_\odot$. This component is
consistent with a late-F subgiant and we propose this is the hitherto
undetected companion in this supersoft system.

We detect AE~Aqr and SS~Cyg in the mid-IR at levels significantly
above the extrapolation from the near-IR measurements. In both
sources the mid-IR flux is variable on short timescales which may
prove difficult to reconcile with intrinsic or reprocessed emission
from circumbinary material. The IR detection of AE~Aqr fits in very
well within the overall picture of a propeller ejecting clouds of
relativistic electrons. The synchrotron emission is first seen in IR
before the electrons cool and emit in the radio regime. The mid-IR
detection of SS~Cyg is exciting and puzzling. The base-line flux may
be due to a free-free wind. The IR emitting region is close to the
wind launching region, explaining the variability. Coherent emission
from magnetic flares is another possibility. Follow-up observations
with {\em SIRTF} of these two objects could provide key insights into
the processes at work.

The non-detections at 4.8 and 11.7~\um\ limit the possibilities for
circumbinary material. We find that any significant amount of
circumbinary material must lay within a small opening angle to avoid
significant visual extinction or a large IR flux.  For CB discs to be
important in the evolution of CVs the discs should be large and
massive and characterized by temperatures $\ga$ 1000~K in their inner
regions.  Recent calculations reveal that the evolution of CB discs to
this phase where they would be detectable \citep{taamsd} is longer
than previously estimated.  The existence of less massive and less
evolved CB discs is a possibility left open by our data. Reprocessing
of the binary emission would make evolved CB discs stand out in the
mid IR unless the inner regions of the CB disc are not efficient
reprocessors (because dust is destroyed and/or cannot form) or if the
CB disc is self-shadowed by its inner edge. In both cases the output
from the CB disc is displaced to the far-IR. Sub-mm observations are
urged to search for the cold, optically thin material at large radii.

\section*{Acknowledgements}
Some of the data presented herein were obtained at the W.M. Keck
Observatory, which is operated as a scientific partnership among the
California Institute of Technology, the University of California and
the National Aeronautics and Space Administration. The Observatory was
made possible by the generous financial support of the W.M. Keck
Foundation. The authors wish to recognize and acknowledge the very
significant cultural role and reverence that the summit of Mauna Kea
has always had within the indigenous Hawaiian community. We are most
fortunate to have the opportunity to conduct observations from this
mountain. The authors wish to acknowledge Fred Chaffee, the W.M, Keck
Observatory Director, for contributing part of a director's night to
acquire some of these data.

\end{document}